\documentclass{llncs}
\usepackage{savesym}
\usepackage{graphicx} 
\def\eventb{EventB\xspace}
\def\why3{Why3}
\savesymbol{because}
\usepackage{amsmath}
\usepackage{z-eves}
\restoresymbol{TXF}{because}
\usepackage{soul}
\usepackage{xspace}
\usepackage{xcolor}
\usepackage{comment}
\usepackage{url}
\usepackage{why3lang}

\newcommand{\setlog}{$\{log\}$\xspace}

\def	\lover		{\mathbin{{\dres} \llap{+\!\!}\,}}
\renewcommand{\oplus}{\lover}

\newcommand{\MACHINE}{\textbf{MACHINE}\xspace}
\newcommand{\REFINES}{\textbf{REFINES}\xspace}
\newcommand{\SEES}{\textbf{SEES}\xspace}
\newcommand{\VARIABLES}{\textbf{VARIABLES}\xspace}
\newcommand{\INVARIANTS}{\textbf{INVARIANTS}\xspace}
\newcommand{\EVENTS}{\textbf{EVENTS}\xspace}
\newcommand{\Initialisation}{\textbf{Initialisation}\xspace}
\newcommand{\tab}{\mbox{}\hspace{5mm}}
\newcommand{\Event}{\textbf{Event}\xspace}
\newcommand{\any}{\textbf{any}\xspace}
\newcommand{\then}{\textbf{then}\xspace}
\newcommand{\Begin}{\textbf{begin}\xspace}
\newcommand{\End}{\textbf{end}\xspace}
\renewcommand{\where}{\textbf{where}\xspace}
\newcommand{\END}{\textbf{END}\xspace}
\newcommand{\ordinary}{$\langle$ordinary$\rangle$}

\title{Comparing \eventb, \setlog and Why3 Models of Sparse Sets}
\author{Maximiliano Cristi\'a\inst{1} \and Catherine Dubois\inst{2}}
\institute{Universidad Nacional de Rosario and CIFASIS, Rosario, Argentina
\and ENSIIE, lab. Samovar, \'Evry-Courcouronnes, France}

\date{}

\begin{document}

\maketitle

\begin{abstract}
Many representations for sets are available in programming languages libraries. The paper focuses on sparse sets used, e.g., in some constraint solvers for representing integer variable domains which are finite sets of values, as an alternative to range sequence. We propose in this paper verified implementations of sparse sets, in three deductive formal verification tools, namely \eventb, \setlog and \why3. Furthermore, we draw some comparisons regarding specifications and proofs.
\end{abstract}

\section{Introduction}

Sets are widely used in programs. They are sometimes first-class objects of programming languages, e.g. SETL \cite{SETL} or \setlog \cite{zbMATH07552282},
but more frequently they are data structures provided in libraries. Many different representations are available, depending on the targeted set operations. In this paper, we focus on sparse sets, introduced by Briggs and Torczon in \cite{BriggsT93}, used in different contexts and freely available for different programming languages (Rust, C++ and many others). In particular, 
sparse sets are used in constraint solvers as an alternative to range sequences or bit vectors for implementing domains of integer variables \cite{sparse_set} which are nothing else than mathematical finite sets of integers. Their use in solvers implementations is motivated by -at least- the {two} following {properties}:  searching and removing  an element are constant-time operations---removing requires only two swapping
operations on arrays; sparse sets are cheap to trail and restore, which is a key point when backtracking.

Confidence on constraint solvers using sparse sets can be improved if the algorithms implementing the main operations are formally verified, as it has been done by Ledein and Dubois in \cite{LedeinD2019} for the traditional implementation of domains as range sequences.  Hence, the main contribution of this paper is
a verified implementation of  sparse sets for representing finite sets of integers in \eventb, \setlog and \why3. 
We prove that the implemented operations preserve the invariants and we also prove properties that can be seen as formal foundations of trailing and restoring. As far as we know, this is the first formally verified implementation of sparse sets, whereas it has been done for other representations e.g. \cite{FilliatreL04,LedeinD2019}. All the specifications and proofs can be found here: \url{https://gitlab.com/cdubois/sets2023.git}.

It has been known for decades that there is no silver bullet for software engineering or software development. The best we can do as software engineers is to increase our toolbox as much as possible and use the best available tool in it for the problem at hand. This software engineer practical principle still applies when it comes to formal development, formal methods and formal verification. In our opinion the Formal Methods (FM for short) community should have as much information as possible about the relative advantages and disadvantages of different FM methods and tools. With the intention to shed some light on the ups and downs of different FM, we specified and verified sparse sets with three different FM {techniques}. Then, a second contribution of this paper is a comparison of these FM w.r.t. aspects such as expressiveness, specification analysis and automated proof.

\section{Sparse sets}
We deal here with sets as subsets of natural numbers up to $N-1$, where $N$ is any non null natural number. A sparse set $S$ is represented by two arrays of length $N$  {called} $mapD$ and $domD$ (as in \cite{sparse_set}), and a natural number $sizeD$. The array $mapD$ maps any value {$v \in [0,N-1]$} to its index $ind_v$  in $domD$, the value indexed by $ind_v$ in $domD$  is $v$. The main idea that brings efficiency when removing an element or testing membership is to split $domD$ into two sub-arrays, $domD[0,sizeD-1]$ and $domD[sizeD, N-1]$, containing resp. the elements of $S$ and the elements of $[0,N-1]$ not in $S$. Then, if $S$ is empty, $sizeD$ 
is equal to 0, if $S$ is the full set, then $sizeD$ is $N$.
Checking if an element $i$ belongs to the sparse set $S$ simply  consists in the evaluation of the expression $mapD[i]<sizeD$. Removing an element from the set consists in moving this element 
to $domD[sizeD, N-1]$ (with 2 \textit{swaps} in $mapD$ and $domD$ and decreasing $sizeD$). Binding $S$ to the singleton set $\{v\}$ follows the same idea: moving this element at the first place in $domD$ and assigning the value 1 to $sizeD$.


In our formalizations, we only deal with two operations  consisting in removing an element in a sparse set and bind a sparse set to a singleton set since these two operations are  fundamental when solving constraints. In this context, we may also need to walk through all the elements of a variable domain, it means exploring $domD[0..sizeD-1]$. If minimal and maximal values are required, then they have to be maintained in parallel. This is outside the scope of this work.

\section{\eventb formal development}

In this section we succinctly introduce the \eventb{} formal specification language and with more detail the \eventb{} models for sparse sets.

\subsection{\eventb}
\eventb{} \cite{abrial2010} is a deductive formal method based on set theory and first order logic allowing users to design correct-by-construction systems. It relies on a state-based modeling language in which a model, called a machine, 
is made of a state and  a collection of events allowing for state changes. The state consists of variables constrained by invariants. 
Proof obligations are generated to verify the preservation of invariants by events. A machine may use a -mathematical- context which introduces abstract sets, constants, axioms or theorems. A formal design in \eventb{} starts with an abstract machine which is usually refined several times. Proof obligations are generated to verify the correctness of a refinement step.

An event may have parameters. When its guards are satisfied, its actions, if any, are executed, updating state variables. Actions may be -multiple- deterministic assignments, $x,y:=e, f$, or -multiple- nondeterministic  ones, $x,y :\mid BAP(x,x',y,y')$ where $BAP$ is called a Before-After Predicate relating current ($x$, $y$) and  next ($x'$, $y'$) values of state variables $x$ and $y$.
In the latter case, $x$ and $y$ are assigned arbitrary values satisfying the BAP predicate. When using such a non-deterministic form of assignment, a feasibility proof obligation is generated in order to check that there exist values for $x'$ and $y'$ such that $BAP(x,x',y,y')$ holds when the invariants and guards hold. Furthermore when this kind of action is used and refined, the concrete action updating $x$ and $y$ is required to assign them values which satisfy the BAP predicate.   

In the following, we use Rodin, an Eclipse based IDE for \eventb{} project management, model edition, refinement and proof, automatic proof obligations generation, model animation and code generation. Rodin supports automatic and interactive provers \cite{DeharbeFGV14}. In this work we used the standard provers (AtelierB provers) and also the SMT solvers VeriT, CVC3  and CVC4. More details about \eventb{} and Rodin can be found in \cite{abrial2010} and \cite{rodin2010}.

\subsection{\eventb formalization}
The formalization is made of six components, i.e. two contexts, a machine and three refinements. Context $Ctx$ introduces the bound $N$ as a non-zero natural number and context $Ctx1$ extends the latter with helper theorems. The high level machine gives the abstract specification. This model contains a state composed of a finite set $D$, constrained to be a subset of the (integer) range $0..N-1$, and two events, to remove an element from $D$ or set $D$ as a singleton set (see Fig. \ref{fig:DomainB} in which $bind$ is removed for lack of space). 


\begin{figure}
\noindent
\MACHINE Domain \\
\SEES Ctx \\
\VARIABLES D \\
\INVARIANTS inv1: $D \subseteq 0 \upto N-1$ \\
\EVENTS \\
\Initialisation \Begin act1: $D := 0\upto{}N-1$ \End \\
\Event remove \ordinary $\defs$ \\
\tab \any v \where grd1: $v \in{} D$ \then act1: $D := D \setminus \{v\}$ \End \\
\END
    \caption{\eventb{} abstract specification, the Domain machine}
    \label{fig:DomainB}
\end{figure}

The first refinement (see Fig.\ref{fig:SparseSetsB1}) 
introduces the representation of the domain as a sparse set, i.e. two arrays $mapD$ and $domD$ modeled as total functions  and also the variable $sizeD$ which is a natural number in the range $0..N$.  Invariants $inv4$ and $inv5$ {constrain $mapD$ and $domD$} to be inverse functions of each other. 
The gluing invariant $inv6$ relates the states between the concrete and {former} abstract machines. So the set $domD[0..sizeD-1]$ containing the {elements of the} subarray from 0 to $sizeD-1$ is exactly the set $D$. 

Theorem $inv7$ is introduced to ease some interactive proofs, it is  proved as a consequence of the previous formulas ($inv1$ to $inv6$). 
It follows directly from a theorem of $Ctx1$ whose statement is $inv7$ where $domD$ and $mapD$ are universally quantified. Theorem $inv8$, also used in an interactive proof, and automatically proved by CVC3, states that $domD$ is an injective function.

Variables $mapD$ and $domD$ are both set initially to the identity function on $0..N-1$ and $sizeD$ to $N$. So invariants are satisfied at the initial state.  Machine \textit{SparseSets\_ref1} refines the events of the initial machine by non deterministic events. So here the  $remove$ event assigns the three state variables with values that satisfy invariants and also such that $sizeD$ strictly decreases and removed elements in $domD$ are kept at the same place (properties in bold font). Event $bind$ follows the same pattern (again not shown here).

\begin{figure}
\noindent
\MACHINE SparseSets\_ref1 \\
\REFINES Domain \\
\SEES Ctx1 \\
\VARIABLES domD mapD sizeD \\
\INVARIANTS \\
\tab inv1: $domD \in 0 \upto N - 1 \fun 0 \upto N - 1$ \\
\tab inv2: $mapD \in 0 \upto N - 1 \fun 0 \upto N - 1$ \\
\tab inv3: $sizeD \in 0 \upto N$ \\
\tab inv4: $domD; mapD = \id_{0 \upto N - 1}$ \\
\tab inv5: $mapD; domD = \id_{0 \upto N - 1}$ \\
\tab inv6: $domD[0 \upto sizeD - 1] = D$ \\
\tab inv7: $\langle$ theorem $\rangle$ \\
\tab\tab $\forall x, v \cdot x \in 0 \upto N - 1 \land v \in 0 \upto N - 1 \implies (mapD(v) = x \iff domD(x) = v)$ \\
\tab inv8: $\langle$ theorem $\rangle$ 
$domD \in 0 \upto N - 1 \inj 0\upto N - 1$ \\
\EVENTS \\
\Initialisation \\
\tab\tab act1: $mapD,domD := \id_{0 \upto N - 1}, \id_{0 \upto N - 1}$\\
\tab\tab act2: $sizeD := N$ \\
\Event remove \ordinary $\defs$ \textbf{refines} remove \\
\tab \any v \\
\tab \where grd1: $v \in 0 \upto N - 1 \land{}$ grd2: $0 < sizeD \land{}$ grd3: $mapD(v) < sizeD$ \\
\tab \then act1: $mapD, domD, sizeD :\mid$ \\
\tab\tab $(domD' \in 0 \upto N - 1 \fun 0 \upto N - 1 \land mapD' \in 0 \upto N - 1 \fun 0 \upto N - 1$ \\
\tab\tab\ ${}\land domD' ; mapD' = \id_{0 \upto N - 1} \land mapD' ; domD' = \id_{0 \upto N - 1}$ \\
\tab\tab\ ${}\land domD'[0 \upto sizeD' - 1] = domD[0 \upto sizeD - 1] \setminus \{v\} \land {\bf sizeD' < sizeD}$ \\
\tab\tab\ ${}\land {\bf (sizeD \upto N - 1) \dres domD' = (sizeD \upto N - 1) \dres DomD}$ \\
\End
       \caption{\eventb{} first refinement}
    \label{fig:SparseSetsB1}
\end{figure}


The second refinement has the same state than the previous refinement (see Fig. \ref{fig:SparseSetsB2}). Its events implement the operations using the new state variables. It is a straightforward translation of the algorithms described in \cite{sparse_set}.

The only reason to have introduced the intermediate model  
\textit{SparseSets\_ref1} is to express the properties written in bold font and thus generate, in the next refinement, proof obligations which, when discharged, will not only ensure that the events refined in Fig. \ref{fig:SparseSetsB2} preserve the invariants $inv1$, $inv2$ \ldots $inv6$ but also the local properties regarding $sizeD$ and $domD[sizeD..N-1]$ (SIM proof obligations).

\begin{figure}
\noindent
\MACHINE SparseSets\_ref2 \\
\REFINES SparseSets\_ref1 \\
\SEES Ctx1  \\
\VARIABLES domD mapD sizeD \\
\EVENTS \\
\Initialisation \\
\tab\tab act1: $mapD,domD := \id_{0 \upto N - 1}, \id_{0 \upto N - 1}$ \\
\tab\tab act2: $sizeD := N$ \\
\Event remove \ordinary $\defs$ \textbf{refines} remove \\
\tab \any v \\
\tab \where grd1: $v \in 0 \upto N - 1 \land{}$ grd2: $0 < sizeD \land{}$ grd3: $mapD(v) < sizeD$ \\
\tab \then \\
\tab\tab act1: $domD := domD \oplus \{mapD(v) \mapsto domD(sizeD - 1), sizeD - 1 \mapsto v\}$ \\
\tab\tab act2: $mapD := mapD \oplus \{v \mapsto sizeD - 1, domD(sizeD - 1) \mapsto mapD(v)\}$ \\
\tab\tab act3: $sizeD := sizeD - 1$ \\
\End
       \caption{\eventb{} second refinement}
    \label{fig:SparseSetsB2}
\end{figure}


The feasibility (FIS) proof obligations generated by the non-deterministic events of \textit{SparseSets\_ref1} require to prove that there exist values such that the BAP predicate holds. We can prove it using the  new values of $domD$, $mapD$ and $sizeD$ specified in the last refinement as witnesses. The simulation (SIM) proof obligations generated by  events of \textit{SparseSets\_ref2} require to prove that the latter values again satisfy the BAP predicate used in \textit{SparseSets\_ref1}. In order not to do these -interactive- proofs twice, we generalize them and prove them as theorems of the context. Thus to discharge the FIS and SIM proof obligations, we only have to instanciate these theorems to provide a proof. 

A last algorithmic refinement, omitted here, refines the $remove$ event in two events, $removeLastButOne$ and $removeLast$. The former differs from $remove$ only by its more restrictive guard; the latter is dedicated to the case where the element with index $sizeD-1$ in $domD$ is removed thus avoiding the unnecessary swapping.


\section{\label{setlog}\setlog formal development}
In this section we briefly present the \setlog tool and how we used it to encode the \eventb model of sparse sets.

\subsection{\setlog}
\setlog is a constraint logic programming (CLP) language and satisfiability solver where sets and binary relations are first-class citizens \cite{setlog,Dovier00,DBLP:journals/jar/CristiaR20}. The tool implements several decision procedures for expressive fragments of set theory and set relation algebra including cardinality constraints \cite{DBLP:journals/tplp/CristiaR23}, restricted universal quantifiers \cite{DBLP:journals/corr/abs-2208-03518}, set-builder notation \cite{DBLP:journals/jar/CristiaR21a} and integer intervals \cite{DBLP:journals/corr/abs-2105-03005}.  In previous works \setlog has been satisfactory tested against some known case studies \cite{DBLP:journals/jar/CristiaR21,DBLP:journals/jar/CristiaR21b,Cristia2023}.

\setlog code enjoys the \emph{formula-program duality}. This means that \setlog code can behave as both a formula and a program. When seen as a formula, it can be used as a specification on which verification conditions can be (sometimes automatically) proved. When seen as a program, it can be used as a (less efficient) regular program. Due to the formula-program duality, a piece of \setlog code is sometimes called \emph{forgram}---a portmanteau word resulting from combining \emph{for}mula with prog\emph{gram}.

\subsection{\setlog formalization}
 The \setlog formalization presented in this paper is the result of translating the \eventb abstract specification (i.e., Fig. \ref{fig:DomainB}) and the second refinement (i.e. Fig. \ref{fig:SparseSetsB2}). Both \eventb models can be easily translated into \setlog by using the (still under development) state machine specification language (SMSL) defined on top of \setlog 
 (see Fig. \ref{fig:setlog1} and \ref{fig:setlog2}) \cite{Rossi00}. The notions of context and refinement are not available in {SMSL}. For this reason, refinements introduced in the \eventb model have to be manually encoded in \setlog. The context is encoded simply as an axiom. In order to ensure that the \setlog code verifies the properties highlighted in bold in Fig. \ref{fig:SparseSetsB1} as well as the gluing invariant (i.e., $inv6$), a few user-defined verification conditions are introduced as theorems. Since the first \eventb refinement is introduced to express the properties written in bold, its events have not been encoded in \setlog.

\begin{figure}
    \centering
\begin{verbatim}
parameters([N,I]).
variables([D,DomD,MapD,SizeD]).

axiom(axm1).
axm1(N) :- 1 =< N.

axiom(axm2).
axm2(N,I) :- M is N - 1 & id(int(0,M),I).

invariant(inv11).
inv11(DomD) :- pfun(DomD).

n_inv11(DomD) :- neg(  pfun(DomD)  ).

invariant(inv12).
inv12(N,DomD) :- N1 is N - 1 & dom(DomD,int(0,N1)).

invariant(inv13).
inv13(N,DomD) :- N1 is N - 1 & ran(DomD,R) & subset(R,int(0,N1)).
  
invariant(inv4).
inv4(N,I,DomD,MapD) :- axm2(N,I) & comppf(DomD,MapD,I).

inv6(D,DomD,SizeD) :-
  S is SizeD - 1 &
  foreach([X,Y] in DomD, X in int(0,S) implies Y in D) &
  foreach(X in D, exists([A,B] in DomD, A in int(0,S) & B = X)).

inv7(MapD,DomD) :-
  foreach([[V,Y1] in MapD, [X,Y2] in DomD],
    (Y1 = X implies Y2 = V) & (Y2 = V implies Y1 = X)  ).
  
theorem(inv7_th).
inv7_th(N,MapD,DomD) :-
  neg(inv4(N,I,DomD,MapD) & inv5(N,I,DomD,MapD) implies inv7(MapD,DomD)).
\end{verbatim}
\caption{Some representative axioms and invariants of the \setlog forgram}
\label{fig:setlog1}
\end{figure}

Figures \ref{fig:setlog1} and \ref{fig:setlog2} list only representative parts of the \setlog forgram.
We tried to use the same identifiers as for the \eventb models as much as possible. In this way, for example, the invariant labeled as $inv6$ in the  \textit{SparseSets\_ref1} machine (Fig. \ref{fig:SparseSetsB1}), is named \verb+inv6+ in the \setlog forgram. The name of variables in \setlog cannot fully complain with those used in the \eventb models because \setlog requires all variables to begin with a capital letter. So, for example, $domD$ in the \textit{SparseSets\_ref1} machine becomes \verb+DomD+ in \setlog.

As can be seen in Fig. \ref{fig:setlog1}, the state machine specification language defined on top of \setlog allows for the declaration of parameters (similar to \eventb context constants), state variables, axioms (similar to \eventb context axioms) and invariants. Parameter \verb+I+ is used to compute the identity relation on the integer interval $[0,N-1]$ as shown in axiom \verb+axm2+, which in turn is used in invariant \verb+inv4+. As \setlog is a CLP language implemented on top of Prolog, it inherits many of Prolog's features. In particular, integer expressions are evaluated by means of the \verb+is+ predicate. Along the same lines, all set operators are implemented in \setlog as constraints. For example, \verb+id(A,R)+ is true when \verb+R+ is the identity relation on the set \verb+A+. The term \verb+int(0,M)+ corresponds to the integer interval $[0,M]$.

Invariants named \verb+inv11+, \verb+inv12+ and \verb+inv13+ correspond to invariant $inv1$ of the \textit{SparseSets\_ref1} machine. Splitting invariants in smaller pieces, is a good practice when using \setlog as a prover because it increases the chances of automated proofs. \verb+n_inv11+ implements the negation of invariant \verb+inv11+. \setlog does not automatically compute the negation of user-defined predicates. As a user-defined predicate can contain existential variables, its negation could involve introducing universal quantifiers which fall outside \setlog's decision procedures. Then, users are responsible for ensuring that all predicates are safe.

In invariant \verb+inv6+ we can see the \verb+foreach+ constraint. This constraint implements the notion of \emph{restricted universal quantifier} (RUQ). That is, for some \setlog formula $\phi$ and set \verb+A+, \texttt{foreach(X in A, $\phi($X$)$)} corresponds to $\forall X.(X \in A \implies \phi(X))$. In a \verb+foreach+ constraint it is possible to quantify over binary relations, as is the case of \verb+inv6+. Hence, we have a quantified ordered pair (\verb+[X,Y]+), rather than just a variable. Likewise, \setlog offers the \verb+exists+ constraint implementing the notion of \emph{restricted existential quantifier} (REQ). The important point about REQ and RUQ is not only their expressiveness but the fact that there is a decision procedure involving them \cite{DBLP:journals/corr/abs-2208-03518}. In \verb+inv6+ these constraints are used to state a double set inclusion equivalent to the \eventb formula $domD[0 .. sizeD - 1] = D$. If the user is not convinced or unsure about the validity of this equivalence (s)he can use \setlog itself to prove it.

Note that \verb+inv7+ is not declared as an invariant because in Fig. \ref{fig:SparseSetsB1} it is a theorem that can be deduced from previous invariants. 
Therefore, we introduce it as a simple predicate but then we declare a theorem whose conclusion is \verb+inv7+. Later, \setlog will include \verb+inv7_th+ as a proof obligation and will attempt to discharge it. Given that \setlog is a satisfiability solver, if $\Phi$ is intended to be a theorem then we ask it to prove the unsatisfiability of $\lnot\Phi$.

\begin{figure}
\centering
\begin{verbatim}
operation(remove).
remove(N,SizeD,MapD,DomD,V,SizeD_,MapD_,DomD_) :-
  M is N - 1 & V in int(0,M) & 0 < SizeD & S is SizeD - 1 &
  MapD = {[V,Y1],[Y2,Y4] / MapD1} & disj({[V,Y1],[Y2,Y4]},MapD1) &
  Y1 < SizeD  &
  DomD = {[S,Y2],[Y1,Y5] / DomD1} & disj({[S,Y2],[Y1,Y5]},DomD1) &
  DomD_ = {[S,V],[Y1,Y2] / DomD1} &
  MapD_ = {[V,S],[Y2,Y1] / MapD1} &
  SizeD_ = S.
theorem(remove_pi_inv6).
remove_pi_invr6(N,SizeD,MapD,DomD,V,SizeD_,MapD_,DomD_) :-
  inv7(MapD,DomD) &
  neg(    inv6(D,DomD,SizeD) &
          remove(V,D,D_) &
          remove(N,SizeD,MapD,DomD,V,SizeD_,MapD_,DomD_)
              implies inv6(D_,DomD_,SizeD_)    ).
theorem(remove_b2).
remove_b2(N,SizeD,MapD,DomD,V,SizeD_,MapD_,DomD_) :-
  neg(    N1 is N - 1 &
          remove(N,SizeD,MapD,DomD,V,SizeD_,MapD_,DomD_) &
          fimg(int(SizeD,N1),DomD_,D)
              implies fimg(int(SizeD,N1),DomD,D)  ).
\end{verbatim}
\caption{The \texttt{remove} operation and some user-defined proof obligations}
\label{fig:setlog2}
\end{figure}

Moving into in Fig. \ref{fig:setlog2} we can see the encoding of the $remove$ operation specified in the \textit{SparseSets\_ref2} machine of Fig. \ref{fig:SparseSetsB2}, along with two user-defined proof obligations. In \setlog, there is no global state so state variables have to be included as explicit arguments of clauses representing operations. Next-state variables are denoted by decorating the base name with an underscore character (e.g., \verb+SizeD_+ corresponds to the value  of \verb+SizeD+ in the next state). Another important difference between the \eventb and the \setlog specifications is that in the latter we can use \emph{set unification} to implement function application. For instance, \verb+DomD = {[S,Y2],[Y1,Y5] / DomD1}+ is equivalent to the \eventb predicate: $\exists y_2, y_5, domD_1. (domD = \{sizeD - 1 \mapsto y_2, y_1 \mapsto y_5\} \cup domD_1)$, where $y_1 = mapD(v)$ (due to the previous set unification). The not-membership constraints following the equality constraint prevent \setlog to generate repeated solutions. Hence, when \verb+remove+ is called with some set term in its fourth argument, this term is unified with \verb+{[S,Y2],[Y1,Y5] / DomD1}+. If the unification succeeds, then the images of \verb+S+ and \verb+Y1+ are available.

As said before, some user-defined proof obligations are introduced as theorems to ensure that the \setlog forgram verifies the gluing invariant (i.e., $inv6$) and the properties written in bold in machine \textit{SparseSets\_ref1}. Precisely, theorem \verb+remove_pi_inv6+ states that if \verb+inv6+ holds and  \verb+remove+ and its abstract version (not shown in the paper) are executed, then \verb+inv6+ holds in the next state.\footnote{\texttt{remove} and its abstract version can be distinguished by their arities.}

Likewise, theorem \verb+remove_b2+ ensures that the second property written in bold in machine \textit{SparseSets\_ref1} is indeed a property of the \setlog forgram. As can be seen, the theorem states that if \verb+remove+ is executed and the functional image\footnote{\texttt{fimg} is a user-defined \setlog predicate computing the relational image through a function---\texttt{fimg} stands for \emph{functional image}.} of the interval from \verb+SizeD+ up to \verb+N-1+ through \verb+DomD_+ is \verb+D+, then it must coincide with the functional image of the same interval but through \verb+DomD+.

Once the specification is ready, we can call the verification condition generator (VCG) and run the verification conditions (VC) so generated:
\begin{verbatim}
{log}=> vcg('sp.pl') & consult('sp-vc.pl') & check_vcs_sp.
\end{verbatim}
VCs include the satisfiability of the conjunction of all axioms, the satisfiability of each operation and preservation lemmas for each and every operation and invariant. The last command above will attempt to automatically discharge every VC. Part of the output is as follows:
\begin{verbatim}
Checking remove_is_sat ... OK
Checking remove_pi_inv11 ... ERROR
\end{verbatim}
An \verb+ERROR+ answer means that, for some reason, \setlog is unable to discharge the VC. Most of the times this is due to some missing hypothesis which, in turn, is due to the way the VCG generates the VCs. Briefly, when it comes to invariance lemmas, the VCG generates them with the minimum number of hypothesis. So, for instance, the invariance lemma named \verb+remove_pi_inv11+ is as follows:
\begin{verbatim}
neg(  inv11(DomD) &
      remove(N,SizeD,MapD,DomD,V,SizeD_,MapD_,DomD_) implies
      inv11(DomD_)  ).
\end{verbatim}

By including minimum hypothesis, \setlog will have to solve a simpler goal which reduces the possibilities to have a complexity explosion. If the hypothesis is not enough, the \verb+findh+ command can be used to find potential missing hypothesis.
In this way, users can edit the VC file, add the missing hypothesis and run the VC again. If more hypotheses are still missing, the process can be executed until the proof is done---or the complexity explosion cannot be avoided.

\setlog discharges all the VC generated by the VCG for the present forgram.

\section{Why3 formal development}

In this section we briefly introduce the \why3 platform and describe with some details our specification of sparse sets.

\subsection{\why3}
Why3 \cite{FilliatreP13} is a platform for deductive program verification providing 
a  language for specification and programming, called WhyML, and relies on external automated and interactive theorem provers, to discharge verification conditions. In the context of this paper, we  used Why3 with the SMT provers CVC4 and Z3.

Proof tactics are also provided, making \why3 a proof environment  close to the one of Rodin for interactive proofs. \why3  supports modular verification.

WhyML allows the user to write functional or imperative programs featuring polymorphism, algebraic data types, pattern-matching, exceptions, references, arrays, etc. These programs can be annotated by contracts and assertions and thus verified. User-defined types with invariants can be introduced, the invariants are verified at the function call boundaries. Furthermore to prevent logical inconsistencies, \why3 generates a verification condition to show the existence of at least one value satisfying the invariant. To help the verification, a witness is explicitly given by the user (see the  \texttt{by} clause in Fig. \ref{fig:whymlTypes}).
The \texttt{old} and \texttt{at} operators can be used inside post-conditions and assertions to refer to the value of a mutable program variable at some past moment of execution. In particular \texttt{old t} in a function post-condition refers to the value of term \texttt{t} when the function is called.  Programs may also contain ghost variables and ghost code to facilitate specification and verification.
From verified WhyML programs, correct-by-construction OCaml programs (and recently C programs) can be automatically extracted.

\subsection{\why3 formalization}

From the \why3{} library, we use pre-defined theories for integer arithmetic, polymorphic finite sets and arrays. In the latter, we use in particular the \texttt{swap} operation that exchanges two elements in an array and its specification using the \texttt{exchange} predicate.

We first define a record type, \texttt{sparse}, whose mutable fields are a record of type \texttt{sparse\_data} containing the computational elements of a sparse set representation  and a ghost finite set of integer numbers which is the abstract model of the data structure. The type invariant of \texttt{sparse} relates the abstract model with the concrete representation. It is used 
to enforce consistency between them. Invariants enforcing consistency between the two arrays \texttt{mapD} and \texttt{domD} and the bound \texttt{sizeD} are attached to the \texttt{sparse\_data} type: lengths of the arrays is \texttt{n}, contents are belonging to $0..\mbox{\texttt{n}}-1$ and the two arrays are  inverse of each other, \texttt{sized} is in the interval $0..\mbox{\texttt{n}}$. These type definitions and related predicates are shown in Fig. \ref{fig:whymlTypes}.

\begin{figure}
    \centering
    \begin{lstlisting}[language=why3]
predicate dom_ran (a: array int) (n: int) =
  0 <= n && a.length = n && forall i. 0<=i<n -> 0<=a[i]< n 

predicate comp_is_id  (a: array int) (b: array int) (n: int) =
  forall v,i.  (0<=i<n &&  0<=v<n) -> (a[i]=v <-> b[v]=i)

type sparse_data = {n: int; mutable domD: array int;
                    mutable mapD: array int; mutable sizeD: int; }
invariant {dom_ran domD n && dom_ran mapD n &&
           comp_is_id domD mapD n && 0 <= sizeD <= n }
by {n = 0 ; domD = make 0 0 ; mapD = make 0 0; sizeD=0}

type sparse = {mutable data: sparse_data; mutable ghost setD: fset int;}
invariant {subset setD (interval 0 data.n) &&
           forall i:int. 0 <= i < data.n ->
              (i < data.sizeD <-> mem data.domD[i] setD)}
by {data = {n = 0 ; domD = make 0 0 ; mapD = make 0 0; sizeD=0} ; 
    setD=FsetInt.empty }
\end{lstlisting}
\caption{WhyML types for sparse sets}\label{fig:whymlTypes}
\end{figure}
    
Our formalization (see Fig. \ref{fig:whymlFun}, where, again, $bind$ is removed for lack of place) contains three functions, 
\texttt{swap\_sparse\_data}, \texttt{remove\_sparse} and \texttt{bind\_sparse}, which update their arguments. They are the straightforward translation of the algorithms in \cite{sparse_set} in WhyML, except for the supplementary ghost code (the last statement in both \texttt{remove\_sparse} and \texttt{bind\_sparse}) which updates the abstract model  contained in \texttt{a.setD}. Function \texttt{swap\_sparse\_data} is a helper function 
which is called in the other ones. The contract of \texttt{swap\_sparse\_data} makes explicit the modifications of both arrays \texttt{a.mapD} and \texttt{a.domD}, using the \texttt{exchange} predicate defined in the library. Verification conditions for this function concern the conformance of the code to the two post-conditions (trivial as it is ensured by \texttt{swap}) and also the preservation of the invariant attached to the \texttt{sparse\_data} type---i.e. mainly that \texttt{a.mapD} and \texttt{a.DomD} after swapping elements remain inverse from each other.
Both \texttt{remove\_sparse} and \texttt{bind\_sparse} act not only on the two arrays and the bound but also on the ghost part, i.e. the corresponding mathematical set \texttt{a.setD}. Thus the verification conditions here not only concern the structural invariants related to \texttt{mapD}, \texttt{domD} and \texttt{sizeD} but  also the ones deriving from the use of the \texttt{sparse} type, proving the link between the abstract logical view (using finite sets) and the computational one implemented through arrays.

\vspace{-0.5cm}

\begin{figure}
\centering
\begin{lstlisting}[language=why3]
predicate same_end (a : array int) (b : array int) (s : int) (n : int) =
  forall i. s <= i < n -> a[i] = b[i]

let swap_sparse_data (a : sparse_data) (i : int) (j : int) 
requires {0<=i<a.n}
requires {0<=j<a.n}
ensures {exchange (old a.domD) a.domD i j}
ensures {exchange (old a.mapD) a.mapD a.domD[i] a.domD[j]} =
  swap a.domD i j;
  a.mapD[a.domD[i]] <- i;
  a.mapD[a.domD[j]] <- j;

let remove_sparse (v : int) (a : sparse) 
requires {0<=v<a.data.n}
requires {a.data.mapD[v] < a.data.sizeD}
requires {a.data.sizeD > 0}
ensures {old a.data.sizeD > a.data.sizeD}
ensures {same_end a.data.domD (old a.data.domD) (old a.data.sizeD) a.data.n} =
  swap_sparse_data a.data a.data.mapD[v] (a.data.sizeD - 1);
  a.data.sizeD <- a.data.sizeD - 1;
  a.setD <- remove v a.setD
\end{lstlisting}
\caption{WhyML functions for sparse sets}\label{fig:whymlFun}
\end{figure}

Observe that types \texttt{sparse\_data} and \texttt{sparse} correspond to the state and invariants of the \eventb refinements. The abstract specification presented in the first \eventb machine becomes a ghost field in WhyML. The invariant of the \texttt{sparse} type corresponds to the \eventb gluing invariant ($inv6$). A similar transposition happens for the operations. Actions in the \eventb abstract events, i.e. updating the abstract set, appear as ghost code in WhyML.

All proofs are discovered by the automatic provers except for some proof obligations related to the \texttt{remove} function. Nevertheless these proofs are simplified thanks to some \why3{} tactics that inject some hints that can be used by the external provers to finish the proofs.

\section{Comparison and discussion}

Set theory is primitive in \eventb and \setlog whereas Why3 which permits to express other theories, provides a theory for it. Rodin uses provers where set theory is primitive but can also  call external provers such as VeriT, Z3 and  CVC4---where set theory is not primitive. However a big effort has been done to process set theory in VeriT, which is often recognized as allowing significant improvements in proofs \cite{MentreMFA12}. 
Why3 relies entirely on external provers where set theory is not primitive. Conversely, \setlog is a satisfiability solver that can only work with set theory---and linear integer algebra. It is the only 
of the three tools implementing advanced decision procedures for set theory. Likely, this proved to be crucial for \setlog being able to be the only tool that automatically discharged all the VC, although it required a simple hypothesis discovery procedure. It should be a concern the time \setlog needs to discharge all the VC because with more complex models the resolution time might be prohibitive. It worth to be studied ways of avoiding the algorithmic complexity  of the decision procedures implemented in \setlog. Results on Computable Set Theory should be revisited (eg. \cite{DBLP:series/mcs/CantoneOP01}). Why3 and Rodin interactive proofs are not numerous and remain quite simple.

In \eventb, 51 proof obligations were generated for the whole development, around half of them coming from the first refinement. 
37 were proven automatically by the standard provers (AtelierB provers), 18 automatically by SMT provers, mainly VeriT, either directly or  after applying the Rodin lasso allowing for  adding additional,
backup hypotheses  having identifiers in common with
the goal. Only two proof obligations required real human intervention, mainly instantiations of the general theorems introduced in $Ctx1$ or explicit witnesses introduction in the case of feasibility proof obligations.

After working in the way described in Sect. \ref{setlog}, \setlog discharges all the 38 VC generated by the VCG in around 7 minutes.

Why3 makes it possible to apply transformations (e.g. split conjunctions) on a proof goal instead of calling an automatic prover on it. Some of these transformations are very simple, e.g. splitting conjunctions, and can then been applied systematically and automatically. Most of the generated VC in our formalization were proven automatically  thanks to the split transformation. Only two of them about pieces of type invariants, required human interaction to insert some more complex transformations,  e.g a case analysis on indexes in mapD (\texttt{case (i=a\_data.mapD[v]}). At the end, 55 VC were proved by CVC4, except two of them discharged by Z3, in a total amount of time of 30 seconds.

Clearly, all three tools are expressive enough for the problem at hand. However, the \eventb specification is probably the most readable. The  tools permit to express axioms, invariants and automatically generate similar VC. \setlog still needs work to express how two models are linked in terms of abstraction/refinement relations. Writing some key properties proved to be complex in \eventb. Indeed, it was necessary to add a somewhat artificial refinement level for Rodin being able to generate the desired VC linking. These properties can be easily defined by the user in \setlog. However, in Why3 and \eventb,  proof obligations are automatically generated from the specifications, in particular the abstract and concrete models can be naturally linked and the tool automatically generates the corresponding VC. In that regard, Why3 and \eventb are safer than \setlog.

The possibility to count with executable code without much effort enables many lightweight analysis that can be put into practice before attempting complex proofs. In \setlog tool where specification and implementation are described by only one piece of code (cf. forgrams). This tool is not the integration of an interpreter and a prover; the same set of rewrite rules are used to compute and prove. In \eventb/Rodin there is only a specification---later it can be converted into an executable representation if tools such as ProB are used. 
Why3 can execute WhyML programs natively thanks to its interpreter and the \texttt{execute} command.
Furthermore, once the the program is proved to verify the specification, correct-by-construction OCaml and C programs can be automatically extracted. These programs will be orders of magnitude more efficient than the equivalent \setlog forgrams.

\section{Conclusion}
We formally verified the implementation of sparse sets using three formal languages and associated tools, focusing on the operations and correctness properties required by a constraint solver when domains of integer variables are implemented with sparse sets. We compared in particular the several statements of invariants and pre-post properties and their proofs. 
As future work, two directions can be investigated. The first one is to complete the formal developments with other set operations. A second one is to implement and verify, in Why3 or \eventb, a labeling procedure such as the ones used in constraint solvers, it would need to backtrack on the values of some domains, and thus make use of the theorems proven in this paper. {Labeling is native in} \setlog {when the CLP(FD) solver is active.}

\bibliographystyle{abbrv}
\bibliography{sets2023.bib}
\end{document}